\documentclass{elsart}
\usepackage{epsfig}

\begin{document}

\begin{frontmatter}

\title{Directed cycles and related structures in random graphs: II---Dynamic properties}

\author[vcb]{Valmir C. Barbosa\corauthref{vcb1}}
\ead{valmir@cos.ufrj.br}
\author[rdsrs]{Raul Donangelo}
\ead{donangel@if.ufrj.br}
\author[rdsrs]{Sergio R. Souza}
\ead{srsouza@if.ufrj.br}
\corauth[vcb1]{Corresponding author.}

\address[vcb]
{Programa de Engenharia de Sistemas e Computa\c c\~ao, COPPE,\\ 
Universidade Federal do Rio de Janeiro,\\
C.P. 68511, 21941-972 Rio de Janeiro, Brazil}
\address[rdsrs]
{Instituto de F\'\i sica, Universidade Federal do Rio de Janeiro,\\ 
C.P. 68528, 21941-972 Rio de Janeiro, Brazil}

\begin{abstract}
We study directed random graphs (random graphs whose edges are directed) as
they evolve in discrete time by the addition of nodes and edges. For two
distinct evolution strategies, one that forces the graph to a condition of
near acyclicity at all times and another that allows the appearance of
nontrivial directed cycles, we provide analytic and simulation results related
to the distributions of degrees. Within the latter strategy, in particular, we
investigate the appearance and behavior of the strong components that were our
subject in the first part of this study.
\end{abstract}


\end{frontmatter}

\section{Introduction}\label{intr}

In the first part of this study \cite{bds03}, we considered random digraphs on
a fixed set of $n$ nodes with Poisson-distributed in- and out-degrees. For a
wide range of values for $z$ (the mean in- or out-degree of a node), we
investigated the behavior of the strong components of the digraph, which
essentially are maximal subgraphs whose nodes can all be reached from one
another by following the directions of the edges. We contributed analytic and
simulation results to the special cases of cycle components and knots. More
specifically, we demonstrated that cycle components are concentrated near $z=1$
and necessarily small, encompassing a number of nodes proportional to $\ln n$,
while knots tend to occur only sparsely as size-one components for very small
values of $z$, then become practically absent as $z$ is increased, then finally
occur again as a single knot that encompasses nearly all $n$ nodes for larger
values of $z$. The latter transition is sharp and happens roughly at
$z=\ln(2n)$.

In this second part, our aim is to study the behavior of strong components
when the digraph is no longer static, but rather evolves in time. This type of
study was also pioneered by Erd\H{o}s and R\'{e}nyi \cite{er60}, who considered
undirected graphs on a fixed set of nodes that progressively become more and
more interconnected by the random addition of edges between pairs of nodes,
thus leading to a Poisson distribution of node degrees. Recently, though, in an
attempt to model the networks that occur in many areas of interest, considerable
effort has been directed towards studying evolution scenarios in which nodes and
edges may enter and leave the network continually. The reader is referred to the
surveys in \cite{bs03} for details related to several areas. One of the surveys
\cite{br03} highlights the mathematics of a variety of such models (e.g.,
\cite{cf03}), many of which exhibit the power-law (as opposed to Poisson)
distribution of degrees that is characteristic of networks that grow by some
sort of preferential (as opposed to random) attachment of nodes to each other.

Our focus is on the study of digraphs that evolve by the addition of new nodes,
as well as new edges, in discrete time. Also, although we do touch the issue of
preferential attachments briefly at one point, for the most part we follow
\cite{bds03} and concentrate on the case of random connections. Even though some
of the models discussed in the literature (cf.\ \cite{bs03}) can be claimed to
include digraphs---at least inherently---we believe our emphasis on random
connections in the evolution of digraphs covers new ground.

The following notation is common to all the remaining sections. For $t\ge 0$ an
integer, we let $D_t$ denote the digraph at time $t$ and $N_t$, with
$n_t=\vert N_t\vert$, its node set. $D_0$ is assumed to have no nodes, so
$N_0=\emptyset$ and $n_0=0$. Unlike the static case \cite{bds03}, it is now
necessary to consider in- and out-degrees separately. For node $i$, we let
$d_t^+(i)$ denote its in-degree at time $t$ and $d_t^-(i)$ its out-degree, with
expected values $z_t^+(i)$ and $z_t^-(i)$, respectively.

The paper is organized as follows. In Section \ref{acyclic} we consider
evolution scenarios in which the deployment of edges disallows every nontrivial
directed cycle. In Section \ref{cycles} we modify the deployment rule in order
to allow the appearance of arbitrary directed cycles, and with them nontrivial
strong components. Conclusions are given in Section \ref{concl}.

\section{Nearly acyclic evolution}\label{acyclic}

\subsection{Analytic results}\label{acyclic:analytic}

In the first evolution scenario that we consider, at every time step $t>0$ a new
node is added to the digraph, so we have $n_t=t$. We assume that nodes are
numbered consecutively from $1$ as they enter the digraph, so node $i$ is the
node added at time $t=i$. When a node enters the digraph, a random number of
edges is also added, all of them directed away from the newly added node. Except
for the possibility of self-loops (an edge leading from a node to itself is
called a self-loop), the digraph is at all times acyclic (i.e., has no directed
cycles spanning more than one node) and all its strong components have one
single node.

For some fixed $z$, we start by considering the case in which an edge is
deployed from node $i$ to each of nodes $1,\ldots,i$ independently with
probability $z/i$ (provided, of course, that $z\le i$; for $z>i$, it must be
assumed that $i$ is connected to all of $1,\ldots,i$). In this case, for
sufficiently large $i$ and $t\ge i$, the out-degree $d_t^-(i)$ is clearly
Poisson-distributed with mean $z_t^-(i)=z$.

As for the in-degree $d_t^+(i)$, its expected value $z_t^+(i)$ is, for
sufficiently large $i$ and $t\ge i$, given by
\begin{equation}
z_t^+(i)=\sum_{u=i}^t\frac{z}{u}=z(H_t-H_{i-1}),
\end{equation}
where $H_m=\sum_{u=1}^m1/u$ is the $m$th harmonic number, given for large $m$ by
$H_m=\ln m+\gamma$, with $\gamma$ denoting Euler's constant \cite{gkp94}. Thus,
for $i\gg 0$,
\begin{equation}
\label{mean-in-uni}
z_t^+(i)\approx\ln\left(\frac{t}{i}\right)^z.
\end{equation}

In order to discover how $d_t^+(i)$ is distributed, let $P_t(i,k)$ denote the
probability that $d_t^+(i)=k$ for $t\ge i$ and $k\ge 0$. If $t-i+1<k$, then
clearly $P_t(i,k)=0$. Otherwise, for $k\ll t-i$ and $z\ll i,t$, we have
\begin{equation}
\label{prob-in-uni1}
P_t(i,k)\approx\frac{P_t(i,0)}{k!}\left[z_t^+(i)\right]^k,
\end{equation}
as we demonstrate in Appendix \ref{derivation1}. $P_t(i,0)$ can now be
approximated from (\ref{prob-in-uni1}) by summing $P_t(i,k)$ over the
appropriate range of $k$:
\begin{equation}
1
\approx P_t(i,0)\sum_{k=0}^{t-i+1}\frac{1}{k!}\left[z_t^+(i)\right]^k
\approx P_t(i,0)e^{z_t^+(i)},
\end{equation}
hence $P_t(i,0)\approx e^{-z_t^+(i)}$ and (\ref{prob-in-uni1}) can be re-written
as
\begin{equation}
\label{prob-in-uni2}
P_t(i,k)\approx\frac{\left[z_t^+(i)\right]^ke^{-z_t^+(i)}}{k!},
\end{equation}
which is the Poisson distribution with mean $z_t^+(i)$.

Denoting the overall in-degree distribution by $P_t(k)$, we see that it can be
estimated by averaging (\ref{prob-in-uni2}) over all nodes, that is,
\begin{equation}
\label{avg-prob-in-uni1}
P_t(k)
\approx\frac{1}{t}\sum_{i=1}^t\frac{\left[z\ln(t/i)\right]^ke^{-z\ln(t/i)}}{k!}
\approx\frac{1}{zk!}\int_{x=0}^{z\ln t}x^ke^{-(1+1/z)x}dx,
\end{equation}
where $x=z\ln(t/i)$ (hence $di=-(t/z)e^{-x/z}dx$). The integrand in
(\ref{avg-prob-in-uni1}) peaks at $x=kz/(z+1)$ with a finite value, so the
integral can be extended to infinity, yielding
\begin{equation}
\label{avg-prob-in-uni2}
P_t(k)
\approx\frac{1}{zk!}\int_{x=0}^{\infty}x^ke^{-(1+1/z)x}dx
=\frac{z^k}{(z+1)^{k+1}}.
\end{equation}

Let us now turn to the case in which node $i$, upon entering the digraph at time
$t=i$, connects out to node $j\in\{1,\ldots,i\}$ with probability proportional
to how many incoming edges $j$ already has (plus $1$, to ensure nonzero
probabilities to start with) while aiming at the same mean out-degree $z$. In
this case, an edge is deployed from $i$ to $j$ with probability
\begin{equation}
\label{prob-var}
z\left(\frac{1+d_{t-1}^+(j)}{t+\sum_{u=1}^td_{t-1}^+(u)}\right),
\end{equation}
with the understanding that $d_{t-1}^+(i)=0$, provided
$z[1+d_{t-1}^+(j)]\le t+\sum_{u=1}^td_{t-1}^+(u)$ (when this does not hold, then
the deployment is assumed to take place with probability $1$). While
(\ref{prob-var}) trivially ensures that the expected value $z_t^-(i)$ of
$d_t^-(i)$ is equal to $z$ for all $t$ that render the expression in
(\ref{prob-var}) a legitimate probability, the underlying distribution of the
out-degree $d_t^-(i)$ is not so easily deduced, although by the results surveyed
in \cite{bs03} its average at time $t$ over all the $t$ nodes already present in
the digraph must be proportional to $k^{-\alpha}$ for some $\alpha>0$.

The same holds for the average distribution $P_t(k)$ of the in-degrees
$d_t^+(i)$. The expected value $z_t^+(i)$ of $d_t^+(i)$, in turn, can be
assessed as follows. By (\ref{prob-var}), the expected probability that node $i$
connects out to node $j$ at time $t=i$ is
\begin{equation}
z\left(\frac{1+z_{t-1}^+(j)}{t+\sum_{u=1}^tz_{t-1}^+(u)}\right).
\end{equation}
Thus, for $t\ge i$, $z_t^+(i)$ obeys the recurrence
\begin{eqnarray}
\label{recurrence-var}
z_t^+(i)
&=&z_{t-1}^+(i)+z\left(\frac{1+z_{t-1}^+(i)}{t+\sum_{u=1}^tz_{t-1}^+(u)}\right)
\nonumber\\
&\approx&z_{t-1}^+(i)+z\left(\frac{1+z_{t-1}^+(i)}{t(z+1)}\right)
\nonumber\\
&=&\left(1+\frac{z}{t(z+1)}\right)z_{t-1}^+(i)+\frac{z}{t(z+1)},
\end{eqnarray}
with $z_t^+(i)=0$ for $t<i$.

It follows from (\ref{recurrence-var}) that
\begin{equation}
\label{mean-in-var1}
z_t^+(i)=\prod_{x=i}^t\left(1+\frac{z}{x(z+1)}\right)-1.
\end{equation}
For $t-i\gg 0$, this solution can be approximated as
\begin{equation}
\label{mean-in-var2}
z_t^+(i)\approx\left(\frac{t}{i}\right)^\frac{z}{z+1}-1,
\end{equation}
as shown in Appendix \ref{derivation2}.

\subsection{Results of simulations}\label{acyclic:simulation}

We have conducted extensive simulations to evaluate the results presented in
Section \ref{acyclic:analytic} on the in-degree distributions when the digraph
evolves in a nearly acyclic fashion. The results we present in this section are
averages of the quantities of interest over a large number of repetitions.

\begin{figure}
\centering
\epsfig{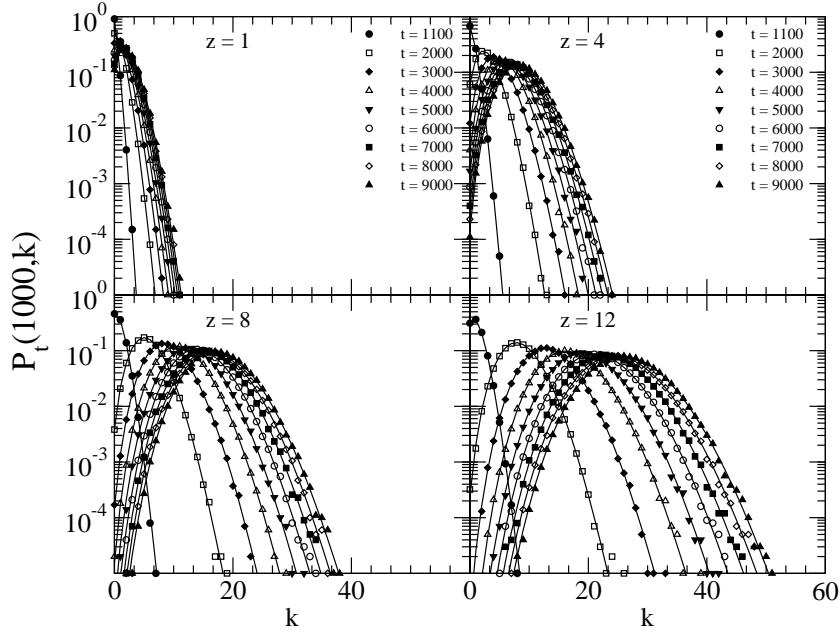}
\caption{Average in-degree distribution in the nearly acyclic case for $i=10^3$
and $z=1,4,8,12$ when edges are deployed uniformly ($10^5$ simulation runs).
Solid plots give the analytic prediction of (\ref{prob-in-uni2}).}
\vspace{0.3in}
\label{pitk-a}
\end{figure}

\begin{figure}
\centering
\epsfig{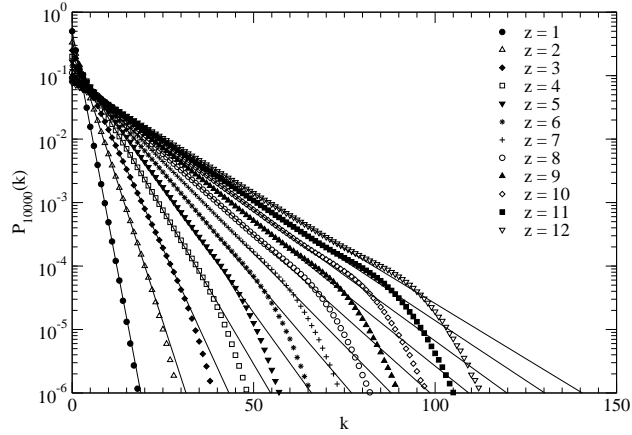}
\caption{Average in-degree distribution in the nearly acyclic case for $t=10^4$
when edges are deployed uniformly ($10^5$ simulation runs). Solid plots give the
analytic prediction of (\ref{avg-prob-in-uni2}).}
\vspace{0.3in}
\label{ptk-a}
\end{figure}

Figures \ref{pitk-a} and \ref{ptk-a} refer to the case in which edges are
deployed uniformly, that is, the Poisson-distributed number of edges that outgo
from each node as it enters the digraph is directed uniformly towards the nodes
already present. Figure \ref{pitk-a} contains plots of $P_t(i,k)$ from both the
simulations and the analytic prediction of (\ref{prob-in-uni2}) for different
values of $z$ and $t$, always with $i=10^3$. As the plots indicate, agreement is
very good throughout.

The plots of Figure \ref{ptk-a} all refer to $P_t(k)$, for which our analytic
prediction is the one in (\ref{avg-prob-in-uni2}). We show data for $t=10^4$ and
several values of $z$. Notice that once again we obtain good agreement between
simulation and analytic prediction, but the two start to separate as $k$
increases. The reason for this is clear: as $k$ increases, the peak of the
integrand of (\ref{avg-prob-in-uni1}), occurring at $x=kz/(z+1)$, shifts
continually to the right, which renders the extension of the integral's upper
limit from the $z\ln t$ of (\ref{avg-prob-in-uni1}) to infinity ever less
justifiable.

\begin{figure}
\centering
\epsfig{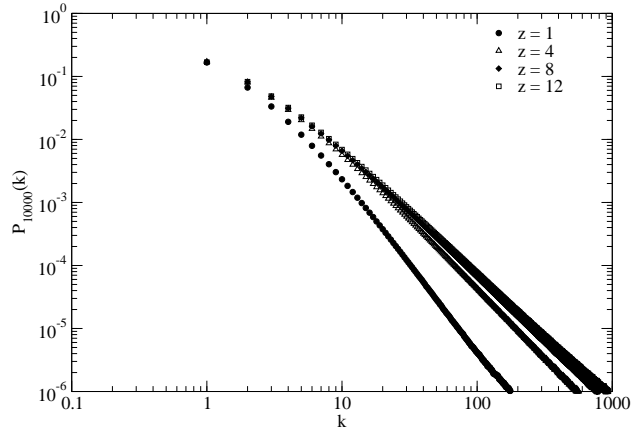}
\caption{Average in-degree distribution in the nearly acyclic case for $t=10^4$
when edges are deployed proportionally to the nodes' in-degrees ($10^5$
simulation runs).}
\vspace{0.3in}
\label{ptk-ap}
\end{figure}

\begin{figure}
\centering
\epsfig{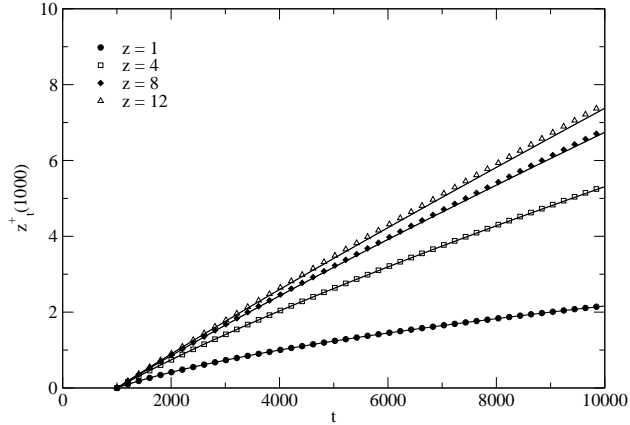}
\caption{Average in-degree in the nearly acyclic case for $i=10^3$ and
$z=1,4,8,12$ when edges are deployed proportionally to the nodes' in-degrees
($10^5$ simulation runs). Solid plots give the analytic prediction of
(\ref{mean-in-var2}).}
\vspace{0.3in}
\label{mean-ap}
\end{figure}

Data for the case of preferential attachment, in which edges get directed
towards nodes already in the digraph with probabilities that are proportional to
how many incoming edges those nodes already have, are given in Figures
\ref{ptk-ap} and \ref{mean-ap}. Figure \ref{ptk-ap} is given for $t=10^4$ and a
set of different $z$ values, and confirms our expectation that in this case
$P_t(k)$ is a power law.

For $i=10^3$ and a few different values of $z$, Figure \ref{mean-ap} contrasts
simulation data for $z_t^+(i)$ with the analytic prediction of
(\ref{mean-in-var2}). Evidently, the two agree very well.

\section{Evolution allowing for cycles}\label{cycles}

\subsection{Analytic results}\label{cycles:analytic}

Our second evolution scenario allows for the appearance of directed cycles other
than self-loops by incorporating a probability, denoted by $p_r$, to control the
replacement of certain edges by certain others. At each time step $t>0$, one of
the following two actions is selected to take place, recalling that the initial
number of nodes is $n_0=0$ and that, as before, nodes are numbered
consecutively from $1$ as they enter the digraph:
\begin{itemize}
\item With probability $p_r$, the number of nodes in the digraph remains the
same, thus $n_t=n_{t-1}$. In addition, a node $i$ for which $d_{t-1}^+(i)>0$ is
selected and one of its incoming edges, chosen randomly with a uniform
distribution, is replaced by a new edge directed away from $i$. Letting
$n_{t-1}^+$ be the number of nodes that have nonzero in-degrees at time $t-1$,
the probability that node $i$ is selected is $1/n_{t-1}^+$, provided its
in-degree is nonzero.
\item With probability $1-p_r$, a new node, say $i$, is added to the digraph,
so $n_t=n_{t-1}+1$. Then a new edge is deployed from $i$ to each of the
digraph's $n_t$ nodes with probability $\min\{z/n_t,1\}$.
\end{itemize}
Readily, for $p_r=0$ this evolution scenario is the one we considered first in
Section \ref{acyclic}. For $p_r>0$, what it does is to induce the appearance of
directed cycles, and consequently of nontrivial strong components.

Under these new rules, $n_t$ is no longer deterministically equal to $t$, nor
is node $i$ necessarily the node added to the digraph at time $t=i$. Instead,
$n_t$ is now a random variable with expected value $(1-p_r)t$, so the expected
time step at which node $i$ is added to the digraph is $t=i/(1-p_r)$, even
though this can happen as early as time $t=i$. Once node $i$ enters the digraph
and receives its Poisson-distributed number of outgoing edges, the process of
edge replacement acts to alter this number of edges randomly, so we still expect
the out-degree $d_t^-(i)$ to be Poisson-distributed with mean $z_t^-(i)=z$.

Analyzing the distribution of the in-degree $d_t^+(i)$, though, is significantly
more complicated and is achieved by setting up a system of finite-difference
equations to describe its behavior. As in Section \ref{acyclic}, $P_t(i,k)$
continues to denote the distribution of in-degrees for node $i$ at time $t$ with
$t\ge i$ and $k\ge 0$. In what follows, we assume that node $i$ is already
present in the digraph at time $t-1$. We do this for the sake of simplicity, and
indicate how to compensate for it when we return to this issue later in this
section.

We start with the $k>0$ case. With probability $1-p_r$, $P_t(i,k)$ is the sum of
two probabilities, each corresponding to one of the following mutually exclusive
events: (i) the node added to the digraph at time $t$ connects out to node $i$;
(ii) the node added to the digraph at time $t$ does not connect out to node $i$.
With probability $p_r$, three other mutually exclusive events must be
considered: (iii) node $i$ is the node selected to have one of its incoming
edges replaced; (iv) node $i$ is not the node selected to have one of its
incoming edges replaced, nor is the replacing edge incoming to $i$; (v) node
$i$ is not the node selected to have one of its incoming edges replaced, but the
replacing edge is incoming to $i$.

Let us assume that, when node $i$ is the node selected to have one of its
incoming edges replaced, the probability that the randomly chosen replacing edge
forms a self-loop, or connects out to a node towards which an edge from $i$
already exists, is negligible. This is certain to hold as the digraph acquires
more nodes, and allows us to conclude the following. In case of event (i) or
(v), the in-degree of node $i$ is increased by $1$ from time $t-1$ to time $t$,
while in cases (ii) and (iv) it remains the same and in case (iii) it is
decreased by $1$. For $k>0$, $P_t(i,k)$ is then such that
\begin{eqnarray}
\label{prob-in}
P_t(i,k)&\approx&(1-p_r)
\left[\min\{z/n_t,1\}P_{t-1}(i,k-1)\right.
\nonumber\\
&&\hspace{0.75in}\mbox{}\left.+\left(1-\min\{z/n_t,1\}\right)P_{t-1}(i,k)\right]
\nonumber\\
&&\mbox{}+p_r\left[\left(\frac{1}{n_{t-1}^+}\right)P_{t-1}(i,k+1)\right.
\nonumber\\
&&\hspace{0.75in}\mbox{}+
\left(1-\frac{1}{n_{t-1}^+}\right)
\left(1-\frac{1}{n_t}\right)P_{t-1}(i,k)
\nonumber\\
&&\hspace{0.75in}\mbox{}+
\left.\left(1-\frac{1}{n_{t-1}^+}\right)
\left(\frac{1}{n_t}\right)P_{t-1}(i,k-1)\right],
\end{eqnarray}
where we have used the fact that $n_t=n_{t-1}$ in the case of edge replacement
at time $t$. For $n_tn_{t-1}^+\gg 1$, and approximating $n_t$ by its expected
value $(1-p_r)t$, (\ref{prob-in}) becomes
\begin{eqnarray}
\label{prob-in-diffeq+}
\lefteqn{P_t(i,k)-P_{t-1}(i,k)
\approx\left(\frac{p_r}{n_{t-1}^+}\right)P_{t-1}(i,k+1)}
\nonumber\\
&&\hspace{1.75in}
-\left((1-p_r)\mu_t+\frac{p_r}{n_{t-1}^+}+\frac{p_r}{(1-p_r)t}\right)
P_{t-1}(i,k)
\nonumber\\
&&\hspace{1.75in}
+\left((1-p_r)\mu_t+\frac{p_r}{(1-p_r)t}\right)P_{t-1}(i,k-1),
\end{eqnarray}
with $\mu_t=\min\left\{z/(1-p_r)t,1\right\}$.

For the case of $k=0$, the only meaningful possibilities are events (ii)--(iv),
hence
\begin{eqnarray}
\label{prob-in-diffeq0}
\lefteqn{P_t(i,0)-P_{t-1}(i,0)
\approx\left(\frac{p_r}{n_{t-1}^+}\right)P_{t-1}(i,1)}
\nonumber\\
&&\hspace{1.75in}
-\left((1-p_r)\mu_t+\frac{p_r}{(1-p_r)t}\right)P_{t-1}(i,0).
\end{eqnarray}

Note in addition that, for $t>0$, it is also possible to approximate $n_t^+$ by
its expected value, that is,
\begin{equation}
\label{expvaluen+}
n_t^+\approx n_t-\sum_{j=1}^tP_t(j,0)
\approx(1-p_r)t-\sum_{j=1}^tP_t(j,0),
\end{equation}
where the summations give the expected number of nodes that have zero in-degree
at time $t$. The system of finite-difference equations given for $k\ge 0$ by
(\ref{prob-in-diffeq+})--(\ref{expvaluen+}) can then be solved numerically for
all $t\ge 1$ and all $i\le t$. The greatest possible in-degree for fixed $i$ and
$t$ is $t-i+1$, corresponding to the case in which node $i$ is added to the
digraph at time step $i$ and acquires a new incoming edge at every subsequent
step, so $k\in\{0,\ldots,t-i+1\}$. All boundary conditions can be set to zero,
and the value of $n_0^+$ is immaterial, so long as it is set to some nonzero
constant.

In order to compensate for the fact that the finite-difference equations being
solved for time $t$ and node $i$ are based on the assumption that node $i$ is in
the digraph at time $t-1$, and also to ensure that a nontrivial solution is
obtained, we may heuristically correct the equations for all $t\ge 1$, all
$i\le t$, and every appropriate $k$, by adding adequate probability terms to
reflect the appearance of edges. If $\pi_t(i)$ is the probability that node $i$
is added to the digraph at time $t$, then we add
$\delta_{1k}\pi_t(i)\min\{z/i,1\}$ to (\ref{prob-in-diffeq+}), where
$\delta_{1k}$ is the Kronecker delta function for $k=1$, and
$\pi_t(i)\left(1-\min\{z/i,1\}\right)$ to (\ref{prob-in-diffeq0}).

As for $\pi_t(i)$, it is given by
\begin{equation}
\pi_t(i)=(1-p_r){{t-1}\choose{i-1}}(1-p_r)^{i-1}p_r^{t-i},
\end{equation}
and thus admits an approximation based on the Poisson distribution, that is,
\begin{equation}
\pi_t(i)\approx(1-p_r)P\left((1-p_r)(t-1),i-1\right),
\end{equation}
where we recall that $P(x,k)$ denotes the Poisson distribution with mean $x$
(cf.\ \cite{bds03}).

\subsection{Results of simulations}\label{cycles:simulation}

\begin{figure}
\centering
\epsfig{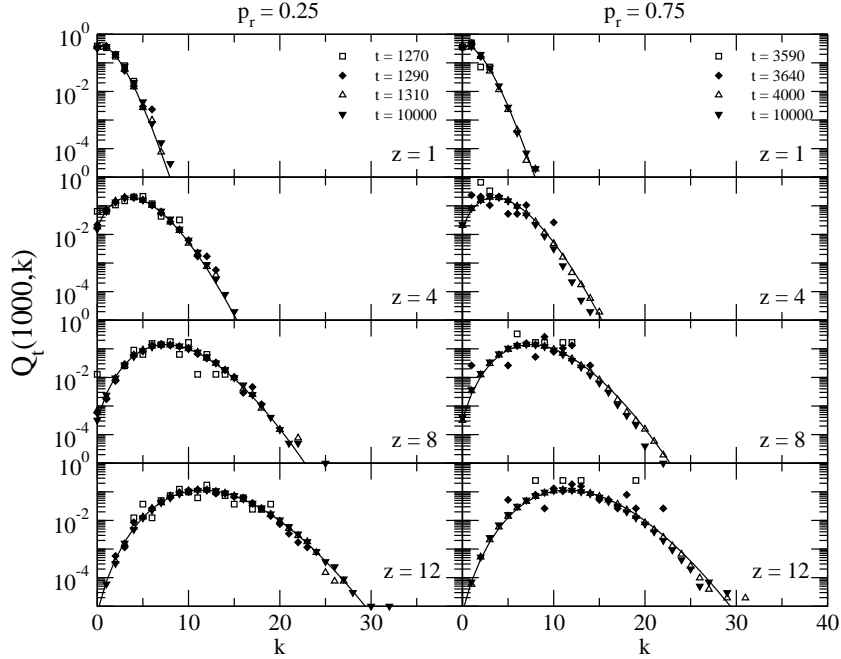}
\caption{Average out-degree distribution under directed-edge replacements for
$i=10^3$, $z=1,4,8,12$, and $p_r=0.25,0.75$ ($10^5$ simulation runs). Solid
plots show the Poisson distribution with mean given by the corresponding value
of $z$.}
\vspace{0.3in}
\label{qitk-r}
\end{figure}

\begin{figure}
\centering
\epsfig{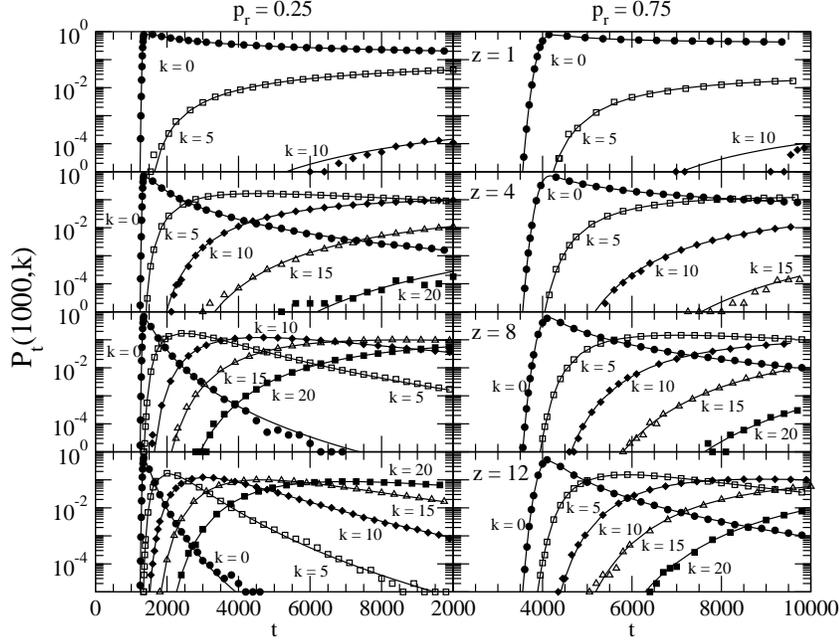}
\caption{Average in-degree distribution under directed-edge replacements for
$i=10^3$, $z=1,4,8,12$, and $p_r=0.25,0.75$ ($10^5$ simulation runs). Panels
placed side by side share the same value of $z$. Solid plots give the solution
of the system of finite-difference equations in
(\ref{prob-in-diffeq+})--(\ref{expvaluen+}).}
\vspace{0.3in}
\label{pitk-r}
\end{figure}

We start with a presentation of simulation data intended to be confronted with
our predictions in Section \ref{cycles:analytic} regarding the distributions of
in- and out-degrees when, during its evolution, the digraph is allowed to have
directed cycles other than self-loops by the action of edge replacements. These
data are presented in Figures \ref{qitk-r} and \ref{pitk-r}.

Figure \ref{qitk-r} contains plots for $p_r=0.25$ and $p_r=0.75$, along with a
few values of $z$ and $t$. Each plot shows simulation data on the distribution
of the out-degree $d_t^-(i)$ of node $i=10^3$. In the figure, we use $Q_t(i,k)$
to denote the probability that $d_t^-(i)=k$, and also show the corresponding
Poisson distribution, $P(z,k)$. As we mentioned earlier, our expectation is for
Poisson-distributed out-degrees, because the number of outgoing edges a node is
given upon entering the digraph is thus distributed, and from there onward all
that the edge-replacement mechanism does is to randomly alter the initial
deployment of edges. We see in Figure \ref{qitk-r} that this is indeed the case
to a very good degree of agreement between $Q_t(i,k)$ and the Poisson
distribution with mean $z$.

Data for the in-degree distribution $P_t(i,k)$ are shown in Figure \ref{pitk-r}
for $i=10^3$ and two values of $p_r$ ($p_r=0.25$ and $p_r=0.75$). All plots are
shown as functions of $t$ for a few different values of $z$ and $k$. For each
combination of $z$ and $k$ values, two plots are given, one for simulation data
and another to depict the solution of the system of finite-difference equations
in (\ref{prob-in-diffeq+})--(\ref{expvaluen+}) that for $k\ge 0$ describes the
behavior of $P_t(i,k)$ for fixed $p_r$ and $z$. It is clear from the figure that
agreement is very good between the two plots in all cases, with the very few
exceptions of some of the higher values of $k$, but such discrepancies are the
result of insufficient statistics for those particular values of $k$, despite
the $10^5$ simulation runs. Agreement is so good, in fact, that solving the
system of equations becomes largely preferable to simulating the system a
sufficiently large number of times, since the latter is unavoidably slower by
several orders of magnitude.

A by-product of our simulations has been the empirical finding that the expected
value of $n_t^+$ is $[z/(z+1)](1-p_r)t$, where we recall that $(1-p_r)t$ is the
expected value of $n_t$. This fact is very useful, since it decouples the
calculation of $n_t^+$ from that of the in-degree distributions of all nodes
(cf.\ (\ref{expvaluen+})). Consequently, it becomes possible to solve the
equations in (\ref{prob-in-diffeq+}) and (\ref{prob-in-diffeq0}) for only a
selected set of nodes of interest.

\begin{figure}
\centering
\epsfig{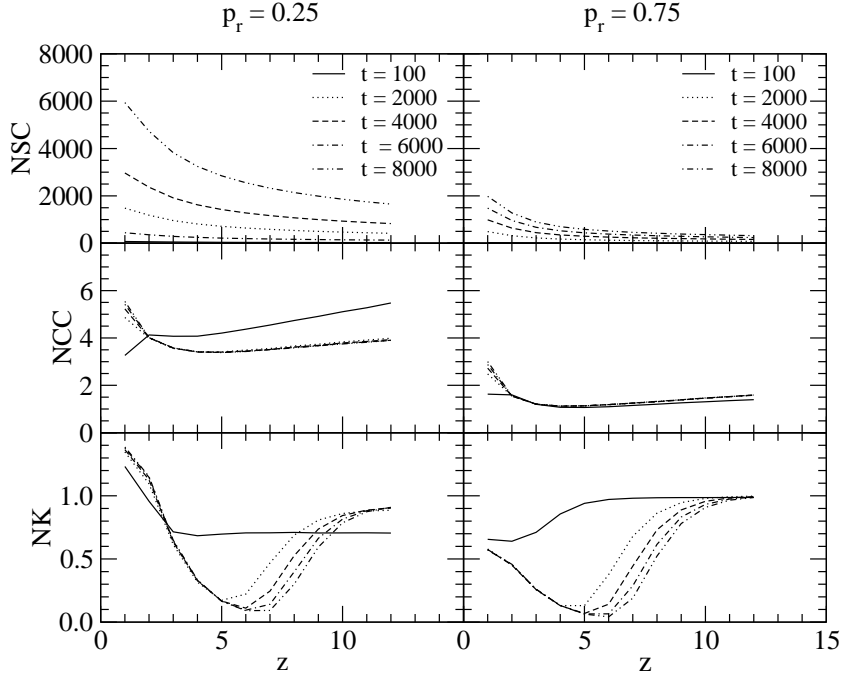}
\caption{Average number of strong components (NSC), of cycle components (NCC),
and of knots (NK) as a function of $z$ for $p_r=0.25,0.75$ ($10^5$ simulation
runs).}
\vspace{0.3in}
\label{nsck}
\end{figure}

\begin{figure}[p]
\centering
\epsfig{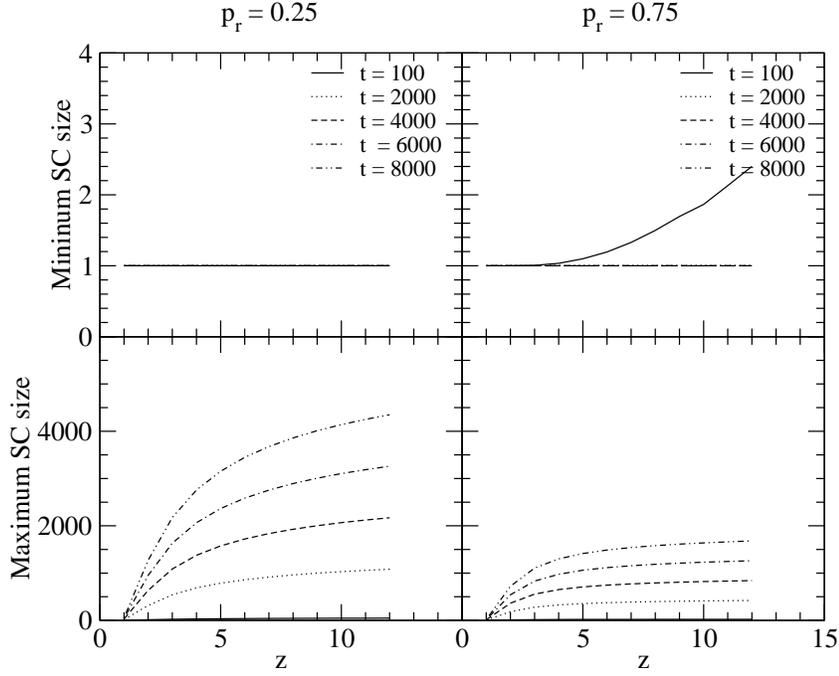}
\caption{Average minimum and maximum sizes of strong components (SC) as a
function of $z$ for $p_r=0.25,0.75$ ($10^5$ simulation runs).}
\vspace{0.3in}
\label{msc}
\end{figure}

\begin{figure}[p]
\centering
\epsfig{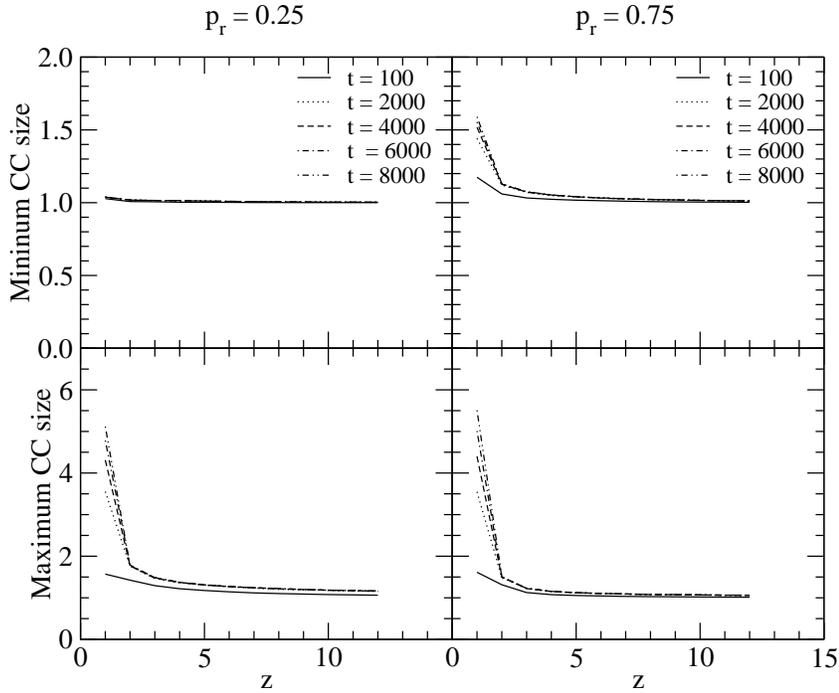}
\caption{Average minimum and maximum sizes of cycle components (CC) as a
function of $z$ for $p_r=0.25,0.75$ ($10^5$ simulation runs).}
\vspace{0.3in}
\label{mcc}
\end{figure}

\begin{figure}
\centering
\epsfig{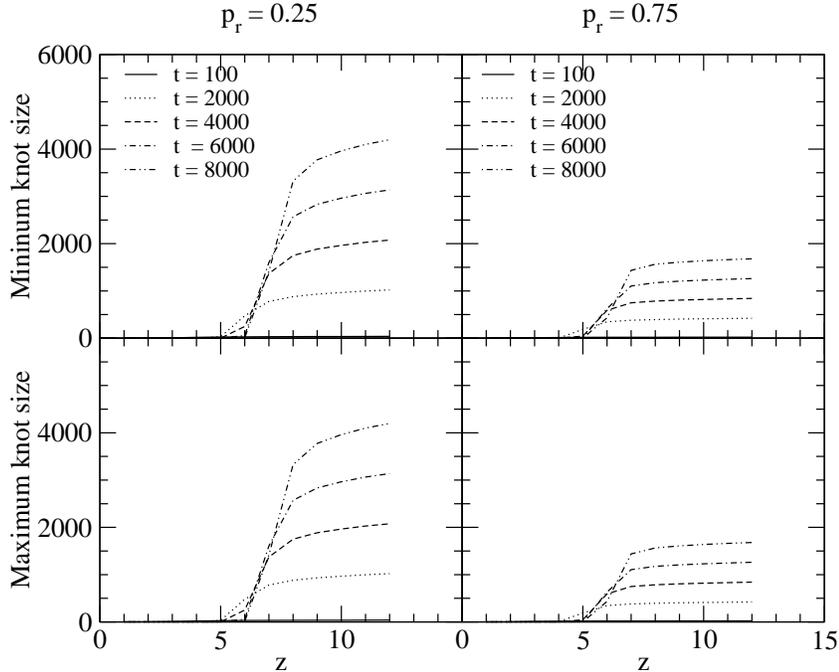}
\caption{Average minimum and maximum sizes of knots as a function of $z$ for
$p_r=0.25,0.75$ ($10^5$ simulation runs).}
\vspace{0.3in}
\label{mk}
\end{figure}

We now present a series of four figures, Figures \ref{nsck}--\ref{mk}, where we
show data related to the behavior of the digraph's strong components, which
under the policy of edge replacement exist in nontrivial form. All the plots
given in these figures are shown as functions of $z$ for $p_r=0.25$, $p_r=0.75$,
and a few relevant $t$ values. The data presented henceforth are then to be
contrasted with the expectation we have from the static case studied in
\cite{bds03}, since the dynamic case is expected to become qualitatively
equivalent to the static case for large enough $t$, so long as $p_r>0$ (the
larger the value of $p_r$, the sooner the equivalence can be observed as $t$
grows). The intuition behind this statement is that, when edges are continually
replaced, the passage of enough time steps is expected to add such a degree of
randomization to the positioning of the edges that it becomes probabilistically
indistinguishable from any positioning that could result if edges were deployed
totally randomly to begin with, as in the static case.

Figure \ref{nsck} is devoted to showing how the number of strong components
behaved during the simulations we conducted, and also in particular the number
of cycle components and of knots. In the static case, the number of strong
components tends to one quickly right past $z=1$ as $z$ is increased and the
giant strong component appears. In the dynamic case we expect a similar effect
as both $z$ and $t$ are increased, that is, we expect the number of strong
components to be dramatically reduced. This is what the figure demonstrates,
particularly for the higher value of $p_r$.

Regarding the number of cycle components, in the static case it peaks at $z=1$
and decreases rapidly to either side, eventually approaching zero. The data in
Figure \ref{nsck} tend to support the expected peak at $z=1$, but even for the
higher value of $p_r$ a nonzero number of cycle components, albeit small, seems
to be sustained as both $z$ and $t$ grow. We will see shortly that this is due
to the presence of cycle components that are self-loops, and presume that it can
be explained by demonstrating that such components have a non-negligible
probability of appearing for small values of $t$, being on the other hand less
likely to be picked for replacement as $t$ increases.

In the static case, knots are expected to comprise one single node and to occur
sparsely for very small values of $z$, then to be altogether absent as $z$ is
increased but still kept below $\ln(2n)$, and then to occur as a single,
all-encompassing knot for $z>\ln(2n)$. In the dynamic case, we expect a similar
behavior, including the effect around the threshold
$z=\ln(2n_t)\approx\ln\left(2(1-p_r)t\right)$. Taking $p_r=0.75$ and $t=8000$,
for example, yields $z=8.3$, which agrees well with what is shown
in Figure \ref{nsck}. But the absence of knots for the intermediate values of
$z$ seems too brief when contrasted with the expectation created by the static
case. Returning to the corresponding data in \cite{bds03}, specifically
Figure 6, we see that our expectation was built on one single simulation run,
which may have been the cause for the seemingly larger intervals of $z$ inside
which knots were totally absent.

These observations are complemented by the data shown in Figures
\ref{msc}--\ref{mk}, where minimum and maximum sizes are shown for the strong
components in general, the cycle components, and the knots, respectively,
observed during the simulations. As shown in Figure \ref{msc}, the minimum size
of a strong component is one, particularly as $t$ grows, while the maximum
approaches $n_t\approx(1-p_r)t$ as $z$ grows away from one, reflecting the
expected appearance of the giant strong component comprising all $n_t$ nodes.

Minimum and maximum sizes observed for the cycle components are as shown in
Figure \ref{mcc}. Aside for a small vicinity near $z=1$, where sizes larger than
one occur, cycle components have size one for all values of $t$. This lends
support to our earlier suspicion that the cycle components still persisting as
both $z$ and $t$ grow are in fact size-one components.

The case of knots is illustrated by the data in Figure \ref{mk}. It shows that
the single knot that appears suddenly roughly around
$z=\ln\left(2(1-p_r)t\right)$ is indeed all-encompassing, as its size tends to
$n_t\approx(1-p_r)t$ as $z$ grows, thus confirming our expectations from the
static case.

\vfill\newpage

\section{Conclusions}\label{concl}

We have in this paper considered random digraphs that grow in discrete time by
the continual addition of new nodes and edges. Within this context, we started
with a study of digraphs that remain nearly acyclic at all times as nodes are
given a Poisson-distributed number of outgoing edges upon entering the digraph.
For two modes of attachment to the nodes already in the digraph, uniform and
preferential, we contributed analytic and simulation results that describe the
digraph's degree distributions.

In the second part of the paper, we allowed edges to be randomly replaced
during the evolution of the digraph, aiming at allowing nontrivial strong
components to appear. In this case, too, we contributed analytic and simulation
results, with emphasis on the introduction of a system of finite-difference
equations for the computation of a node's in-degree distribution at all times.
We finalized by returning to the strong components that were our main subject
in \cite{bds03} and investigated their appearance and behavior along the
evolution of the digraph.

We mention that it is possible to extend our study of the evolution of a digraph
and its strong components to, in principle, all kinds of connection rules and
evolution strategies, particularly those that have special significance within
a certain application area. The ones we studied were ultimately targeted at the
study of strong components, and for this reason the absence or presence of
directed cycles, as well as the conditions for their appearance during the
evolution, have been crucially important. Other guiding principles will
certainly exist as the focus is shifted by some other motivation.

\appendix

\section{Derivation of (\ref{prob-in-uni1})}\label{derivation1}

Note first that, for each subset of $\{i,\ldots,t\}$ comprising $k$ nodes,
$P_t(i,k)$ expresses the probability that an edge exists directed from each of
those $k$ nodes to node $i$, but not from any of the remaining $t-i+1-k$ nodes
of $\{i,\ldots,t\}$. Let $\mathcal{C}$ denote the set of all partitions
$(C,\bar C)$ of $\{i,\ldots,t\}$ such that $C$ contains exactly $k$ nodes. Then
we have
\begin{eqnarray}
\label{app-eq1}
P_t(i,k)
&=&\sum_{(C,\bar C)\in\mathcal{C}}
\prod_{x\in C}\frac{z}{x}
\prod_{x\in\bar C}\left(1-\frac{z}{x}\right)
\nonumber\\
&=&\sum_{(C,\bar C)\in\mathcal{C}}
\prod_{x\in C}\left[1-\left(1-\frac{z}{x}\right)\right]
\prod_{x\in\bar C}\left(1-\frac{z}{x}\right)
\nonumber\\
&=&\sum_{(C,\bar C)\in\mathcal{C}}\sum_{j=0}^k(-1)^{k-j}
\sum_{{B\subseteq C}\atop{\vert B\vert=j}}
\frac{\prod_{x=i}^t\left(1-\frac{z}{x}\right)}
{\prod_{x\in B}\left(1-\frac{z}{x}\right)}.
\end{eqnarray}

Since $P_t(i,0)=\prod_{x=i}^t(1-z/x)$, we can write (\ref{app-eq1}) as
\begin{eqnarray}
\label{app-eq2}
P_t(i,k)
&=&P_t(i,0)\sum_{j=0}^k(-1)^{k-j}
\sum_{(C,\bar C)\in\mathcal{C}}\sum_{{B\subseteq C}\atop{\vert B\vert=j}}
\prod_{x\in B}\left(1+\frac{z}{x-z}\right)
\nonumber\\
&=&P_t(i,0)\sum_{j=0}^k(-1)^{k-j}
{t-i+1-j\choose k-j}
\sum_{{B\subseteq C\cup\bar C}\atop{\vert B\vert=j}}
\prod_{x\in B}\left(1+\frac{z}{x-z}\right)
\nonumber\\
&=&P_t(i,0)\sum_{j=0}^k(-1)^{k-j}
\frac{{t-i+1\choose k}}{{t-i+1\choose j}}{k\choose j}
\sum_{{B\subseteq C\cup\bar C}\atop{\vert B\vert=j}}
\prod_{x\in B}\left(1+\frac{z}{x-z}\right).
\end{eqnarray}

For $k\ll t-i$, in (\ref{app-eq2}) we can use the approximations
\begin{equation}
\frac{{t-i+1\choose k}}{{t-i+1\choose j}}
\approx\frac{j!(t-i+1)^k}{k!(t-i+1)^j}
\end{equation}
and
\begin{eqnarray}
\sum_{{B\subseteq C\cup\bar C}\atop{\vert B\vert=j}}
\prod_{x\in B}\left(1+\frac{z}{x-z}\right)
&\approx&
\frac{1}{j!}
\left[\sum_{x\in C\cup\bar C}\left(1+\frac{z}{x-z}\right)\right]^j
\nonumber\\
&\approx&
\frac{1}{j!}
\left[t-i+1+\ln\left(\frac{t-z}{i-z-1}\right)^z\right]^j,
\end{eqnarray}
which yield
\begin{eqnarray}
P_t(i,k)
&\approx&\frac{P_t(i,0)}{k!}\sum_{j=0}^k{k\choose j}(-t+i-1)^{k-j}
\left[t-i+1+\ln\left(\frac{t-z}{i-z-1}\right)^z\right]^j
\nonumber\\
&=&\frac{P_t(i,0)}{k!}\left[\ln\left(\frac{t-z}{i-z-1}\right)^z\right]^k
\end{eqnarray}
and, for $z\ll i,t$,
\begin{equation}
P_t(i,k)
\approx\frac{P_t(i,0)}{k!}\left[\ln\left(\frac{t}{i}\right)^z\right]^k
\approx\frac{P_t(i,0)}{k!}\left[z_t^+(i)\right]^k,
\end{equation}
by (\ref{mean-in-uni}).

\section{Derivation of (\ref{mean-in-var2})}\label{derivation2}
                                                                                
From (\ref{mean-in-var1}), we have
\begin{eqnarray}
\label{app-eq3}
z_t^+(i)
&=&\prod_{x=i}^t\left(1+\frac{z}{x(z+1)}\right)-1
\nonumber\\
&=&
\sum_{j=0}^{t-i+1}
\left(\frac{z}{z+1}\right)^j\sum_{i\le x_1<\cdots<x_j\le t}
\frac{1}{x_1\cdots x_j}-1.
\end{eqnarray}
For $t-i\gg 0$, in (\ref{app-eq3}) the innermost summation can be approximated
by
\begin{equation}
\frac{1}{j!}\left(\sum_{x=i}^t\frac{1}{x}\right)^j
\approx\frac{1}{j!}\left[\ln\left(\frac{t}{i-1}\right)\right]^j,
\end{equation}
thus yielding, for $i\gg 0$,
\begin{equation}
z_t^+(i)
\approx\sum_{j=0}^{t-i+1}\frac{1}{j!}
\left[\ln\left(\frac{t}{i}\right)^\frac{z}{z+1}\right]^j-1
\approx\left(\frac{t}{i}\right)^\frac{z}{z+1}-1.
\end{equation}

\ack

The authors acknowledge partial support from CNPq, CAPES, the PRONEX initiative
of Brazil's MCT under contracts 41.96.0857.00 and 41.96.0886.00, and a FAPERJ
BBP grant.

\bibliography{rgraph}
\bibliographystyle{elsart-num}

\end{document}